# Interfacial Strain Modulated Correlated Plasmons in La$_{1.85}$Sr$_{0.15}$CuO$_4$ and Their Role in High-temperature Superconductivity

*Xiongfang Liu [1,#], Shengwei Zeng [2,#], Xun Liu [3,#], Kun Han [4], Difan Zhou [1], Chi Sin Tang [5,\*], Ping Yang [5], Mark B. H. Breese [5,6], Chuanbing Cai [1], Ariando Ariando [6,\*], Mi Jiang [3,7,\*], Xinmao Yin [1,\*]*

[1] Shanghai Key Laboratory of High Temperature Superconductors, Department of Physics, Shanghai University, Shanghai 200444, China

[2] Institute of Materials Research and Engineering (IMRE), Agency for Science, Technology and Research (A*STAR), 2 Fusionopolis Way, Innovis #08-03, Singapore 138634, Singapore

[3] School of Physical Science and Technology, Soochow University, Suzhou 215006, China

[4] Information Materials and Intelligent Sensing Laboratory of Anhui Province, Institutes of Physical Science and Information Technology, Anhui University, Hefei 230601, China

[5] Singapore Synchrotron Light Source (SSLS), National University of Singapore, Singapore 117603, Singapore

[6] Department of Physics, Faculty of Science, National University of Singapore, Singapore 117542, Singapore

[7] Jiangsu Key Laboratory of Frontier Material Physics and Devices, Soochow University, Suzhou 215006, China

[#]These authors contributed equally to this work.

*Corresponding author: slscst@nus.edu.sg (C.S.T.); ariando@nus.edu.sg (A.A.); jiangmi@suda.edu.cn (M.J.); yinxinmao@shu.edu.cn (X.Y.)

**Abstract**

High-temperature superconductivity in cuprate materials remains a major challenge in physics due to the complexity of their strongly correlated electronic states. Interfacial strain is a powerful lever for tuning electronic correlations in complex oxides, offering new pathways to control emergent quantum phases. Here, we report the discovery of interfacial strain modulated correlated plasmons observed exclusively in superconducting $La_{1.85}Sr_{0.15}CuO_4$ (LSCO) through spectroscopic ellipsometry. This form of plasmons is absent in the non-superconducting LSCO counterparts. Detailed analysis reveals that these correlated plasmons, arising from the collective excitations within Mott-correlated bands, are driven by long-range electronic correlations in the Cu-O planes. Furthermore, long-range electronic correlations, intricately modulated by interfacial strain, may play a crucial role in the emergence of superconductivity and in tuning the transition temperature. Dynamical cluster approximation (DCA) with quantum Monte Carlo (QMC) calculations of the extended Hubbard model suggest that long-range Coulomb interactions play an important role in LSCO, showing good agreement with our experimental findings. The collective evidence from both the experimental results and theoretical findings provides new insights into the nature of collective excitations and their pivotal role in the emergence of high-temperature superconductivity.

# 1. Introduction

The development of experimental and computational research in the study of strongly-correlated systems has been progressing rapidly. Electronic correlations and magnetic interactions play a crucial role in determining the properties of physical systems, from atoms to molecules, and to condensed-matter phases, where they give rise to diverse phenomena such as metal-to-insulator transitions, giant magnetoresistance, and high-temperature superconductivity [1-4]. Strongly correlated systems manifest themselves over a wide range of energy scales in terms of coherent excitations. High-energy excitations, such as collective spin effect strongly influence the material's physical properties, and set the stage for phenomena observed at lower energies [5]. Meanwhile, low-energy particle coupling is an essential element in mediating electron dynamics of such correlated systems [6]. The intricate interplay between high- and low-energy structures helps govern the macroscopic behavior of these systems from electronic structures to phase transitions [5, 7, 8].

A notable case in point is the cuprate high-temperature superconductors (HTS), where the parent compound is a Mott insulator with antiferromagnetic properties, consisting of both low- and high-energy excitation structures [9, 10]. These important energy-related excitations give rise to the electronic and structural properties and are essential for the formation of Cooper pairing and the opening of the superconducting gap [7, 11-13]. Nevertheless, an unresolved issue at the core of cuprate HTS is a proper minimum model that correctly captures their low-energy properties. The celebrated single-band Hubbard model describes the antibonding band of $Cu\ 3d_{x^2-y^2}$ and $O\ 2p$ orbitals [14, 15], which stems from the simplification of three-orbital Emery model [16]. The understanding of the simplest single-band model remains challenging nonetheless [17], and numerous studies on the impact of nearest-neighbor interaction, in the form of repulsive [18-21] or attractive nature have been proposed [22, 23]. Needless to say, studies concerning the role played by even longer-range interactions remain scarce. Nevertheless, these interactions may be essential for achieving a more comprehensive understanding of the low-energy physics in cuprates. In particular, such interactions at a longer range may play a potentially crucial role in the higher-energy regime associated with plasmon excitations in cuprate high-temperature superconductors, which is the primary focus of this study.

In this work, we employed photon energy-dependent high-resolution spectroscopic ellipsometry to investigate optimal-doped La$_{1.85}$Sr$_{0.15}$CuO$_4$ (LSCO) films grown on various substrates. While conventional (metallic) plasmons can be detected in all our LSCO film samples independent of their superconducting properties, these unique correlated plasmons which are likely associated with high-energy electronic excitations only appear in LSCO films with superconductive features. A series of detailed optical measurements and analyses revealed that different plasmonic states and anomalous optical signals can be attributed to shifts and modifications to the electronic structure that are mediated by the high-energy electronic excitations. Such interactions in turn influence the long-range electronic correlations. Further analysis of the relationship between superconducting transition temperature and interface strain suggests that interfacial strain plays a pivotal role in regulating the transition temperature via tuning the long-range electronic correlations within the Cu–O planes (Figure 1a). Our findings show that long-range electronic correlations are a crucial factor in the formation of superconductivity, thereby providing deeper insights into energy-related excitations and opening new avenues for interpreting the superconductivity mechanism.

## 2. Results and Discussion

### 2.1 Lattice Structure and Characterization Transport measurements

Optimal doped $La_{1.85}Sr_{0.15}CuO_4$ (LSCO) epitaxial films of three different thicknesses (9.0 ± 0.1 nm, 40.0 ± 0.1 nm, 90.0 ± 0.1 nm) have been synthesized on (001)-oriented $LaSrAlO_4$ (LSAO), $LaAlO_3$ (LAO), $SrTiO_3$ (STO) and $(LaAlO_3)_{0.3}(Sr_2TaAlO_6)_{0.7}$ (LSAT) substrates, respectively, with nearly identical in-plane lattice parameters and structure. In which case, LSCO belongs to the $K_2NiF_4$ type structure (space group: I4/mmm) with lattice parameters $a_{bulk}=b_{bulk}=$ 3.777 Å; and $c_{bulk}$=13.226 Å [24, 25].

The representative X-ray diffraction (XRD) $2\theta-\theta$ scans of LSCO/LSAO thin films with varying thicknesses are displayed in Figure 1b. Each LSCO/LSAO film displays sharp (00$l$) Bragg reflections, with no indications of impurity phase orientations. The presence of Laue fringes, indicated by arrows, suggests a highly coherent interface between the film and substrate [26]. Moreover, the spacing of these fringes can be used to confirm that the films possess a flat surface and high quality [27]. These scan results confirm that the films are epitaxially grown along the c-axis with high structural integrity. The detailed structure characterization of the LSCO sample is further confirmed via X-ray reciprocal space mapping (RSM) studies. The RSMs around the $(004)_{HL}$ and $(004)_{KL}$ peaks demonstrate that the LSCO layers align directly beneath the LSAO substrate peak (Figure 1c), indicating fully coherent strain between the LSCO film and LSAO substrate [28]. Furthermore, the RSMs of the $(-109)_{HL}$ and $(019)_{KL}$ peaks reveal that the absolute values of $H$ and $K$ for the LSCO peaks are equal, and that they exhibit the same height as $L$, thereby suggesting a tetragonal symmetry in both the LSCO film and the LSAO substrate. There is also no evidence of tilt in the tetragonal lattice [28]. The XRD and RSM results for LSCO films of variable thicknesses on other substrates are displayed in Figures S1 and S2.

Figure 1d displays the transport measurement for the LSCO/LSAO films, where a sharp superconducting transition is observed in both the ~40.0 and ~90.0 nm films at superconducting transition temperatures ($T_c$) ~21.36 and ~24.95 K, respectively. The results indicate that $T_c$ decreases as decreasing film thickness, eventually vanishing in the ~9.0 nm LSCO thin film. Similar transport data for the other LSCO films (Figure S3) and their $T_c$ are displayed in Figure 1e. The emergence of superconducting properties consistently correlates with increasing LSCO film thickness, while no superconducting transition occurs in the thinner LSCO films [29, 30].

## 2.2 Observation of Correlated Plasmons Excitation

To investigate the superconductive properties and the evolving electronic structures of LSCO with varying thicknesses and substrates, high-resolution spectroscopic ellipsometry (SE) has been employed to examine how the quantitative optical response evolves at these nanometer-thickness LSCO films. Figure 2 shows the real ($\varepsilon_1$) and imaginary ($\varepsilon_2$) components of the dielectric function (where $\varepsilon(\omega)= \varepsilon_1(\omega)+ i\varepsilon_2(\omega)$) of the LSCO/LSAO films with different thickness with the corresponding loss function (LF) spectra ($Im(-1/(\varepsilon(\omega)))$) at 300 K elucidated from the SE measurements. The $\varepsilon_1$ component corresponds to the polarity of the material, while $\varepsilon_2$ indicates its electrical conductivity properties. The characteristic peaks appear in $\varepsilon_2$ spectrum correspond to the optical transitions near the Fermi level ($E_F$) and those in the LF spectrum are associated with optical transitions and plasmon excitations in the system.

The superconducting properties of cuprate superconductors are fundamentally defined by their normal-state electronic structure. LSCO/LSAO exhibits clear metallic behavior, evidenced by a prominent zero-crossing in the $\varepsilon_1$ spectrum at ~0.69 eV (Figure 2a) and a Drude response in $\varepsilon_2$ spectrum (Figure 2b) at 300 K. Meanwhile, the films of ~40.0 nm and ~90.0 nm exhibit higher-intensity Drude responses in $\varepsilon_2$ spectrum compared to the ~9.0 nm LSCO thin film. This indicates that thicker films possess superior metallic properties in the normal state that is consistent with the transport measurements (Figure 1d). Meanwhile, the $\varepsilon_2$ spectra of LSCO films with varying thicknesses possess three principal optical features (Figure 2b) – the Drude response and features *B* and *C*. The Drude response is attributed to intraband transitions near the Fermi level ($E_F$) which contributes to its metallic behavior. Feature *B* (~1.50 eV) is related to ligand-to-metal charge-transfer excitations (O 2*p*→Cu 3*d*) [31, 32]. In contrast, feature *C* (~3.0 eV) can be ascribed to the interband excitations into higher-energy conduction bands [33, 34].

To further analyze the optical properties of the LSCO/LSAO system, the LF spectra of LSCO/LSAO films over the 0.6–4.0 eV energy range were derived, as shown in Figure 2c. Feature *A* at ~0.92 eV in the LF spectrum nearly coincides with the zero-crossing of the $\varepsilon_1$ spectrum, which indicates the presence of a conventional (metallic) plasmon in all LSCO/LSAO samples across different thicknesses [35]. Some disparity between the energy positions of the zero-crossing in the $\varepsilon_1$ spectrum and that of feature *A* in LF spectrum arises from finite free-electron scattering [36, 37]. The positions of the conventional (metallic) plasmons (feature *A*) exhibit noticeable thickness-dependence as it gradually blueshifts from 0.78 eV to 0.92 eV with increasing LSCO films thickness. This is related to changes in the concentration of charge carriers in the normal state (metallic state) [38]. Observations of conventional (metallic) plasmons in similar energy positions have also been reported in other metallic phase cuprates [34-36]. Meanwhile, features *B* (~1.50 eV), and *C* (~3.0 eV) could be further observed in the LF spectra that originate from the charge-transfer excitations and interband transitions, respectively, corresponding to the $\varepsilon_2$ spectra.

A clear distinction of spectral weight (SW) in LF spectra between different thicknesses at ~1.84 eV can be observed from Figure 2c. With this thickness-dependent trend intricately related to the SW and energy-related excitations properties in consideration, a more in-depth analysis of the LF data is conducted and they are plotted in the form of a differential spectra of LF between various temperatures and the 300 K ($\Delta$LF, where $\Delta$LF=LF($\omega$, T)-LF($\omega$, 300 K)), as shown in Figures 2d-f. Compared to the ~9.0 nm non-superconducting LSCO/LSAO film (Figure 2f), the $\Delta$LF spectra show two prominent features below 2.0 eV (features *A* (~0.92 eV) & $A^*$ (~1.84 eV), the shoulder feature *B* at ~1.50 eV is covered) and a smaller feature *C* at ~3.0 eV in the ~90.0 nm and ~40.0 nm superconducting LSCO/LSAO films (Figures 2d-e). The two prominent features *A* & $A^*$, which display a two-fold energy relation and similar temperature dependence, indicate a common origin. However, feature $A^*$ at ~1.84 eV is absent in the $\Delta$LF spectrum of ~9.0 nm non-superconducting LSCO/LSAO film, thereby revealing the shoulder feature *B* (Figure 2f). The weaker feature *C* in each LSCO/LSAO film arises from the temperature-induced energy shift of high-energy interband transitions, reflecting changes in the electronic structure of LSCO. Therefore, feature $A^*$ is the principal difference between superconducting (~90.0 nm and ~40.0 nm) and non-superconducting (~9.0 nm) LSCO/LSAO films. Meanwhile, effective electron number (per Cu atom) $N_{eff}$ is estimated in order to quantify spectral data (Figure S12). The increased $N_{eff}$ in ~90.0 nm and ~40.0 nm superconducting LSCO films on varying substrates also demonstrate the existence of unique electronic excitations at ~ 1.84 eV. Therefore, all the analysis on LF spectra indicate that the SW of ~90.0 nm and ~40.0 nm superconducting LSCO films are different from ~9.0 nm non-superconducting LSCO/LSAO films which is primarily attributed to the additional feature $A^*$ at ~1.84 eV.

The Drude–Lorentz model was employed to fit the loss function (LF) spectra of the LSCO/LSAO film, allowing precise determination of the energy positions and spectral weights (SW) of each feature. The corresponding fitting curves are presented in Figure 3, while the fitting error is shown in Figure S10. Upon further scrutiny of the fitting LF spectra, notice features *A* and $A^*$ which possess a two-fold energy relation located at ~0.92 and ~1.84 eV respectively, in both the ~40.0 and ~90.0 nm LSCO/LSAO films (Figures 3a-b). Intriguingly, this pair of features persists throughout the entire temperature range and that by analyzing the temperature-dependent LF spectra, their energy positions remain fixed throughout the entire temperature range between 77 and 300 K (Figure S5). These observations indicate that the two-fold energy relation of features *A* and $A^*$ unique to the LF spectra is related to the presence of another class of high-energy plasmons in LSCO thick films. Notably, the SW of feature $A^*$ gradually diminishes with decreasing film thickness, disappearing entirely in the ~9.0 nm LSCO/LSAO film, which lacks superconducting properties (Figure 3c). This allows us to deduce that while the unique high-energy plasmon pair is present in LSCO/LSAO films with thicknesses of ~40.0 and ~90.0 nm, it dissipates below a certain LSCO film thickness where it is no longer superconducting.

Similar experimental observations concerning feature *A* and $A^*$ in the LF spectra on LSCO films grows on different substrate can be observed in Figures 3d-f and Figures S6-8. Features $A^*$ which exhibits a two-fold energy relation with feature $A$ only appear in the superconducting compound ~90.0 nm LSCO/LAO, LSCO/STO and LSCO/LSAT thick films, while disappear in the non-superconducting compound ~9.0-nm LSCO/LAO, LSCO/STO and LSCO/LSAT thin films. Based on the behavior of features $A$ and $A^*$, which appear only in the LF spectra and exhibit a two-fold energy relation, their characteristics are consistent with those of correlated plasmons reported in both theoretical and experimental studies [39-41].

## 2.3 The effect of Long-range electronic correlations

Correlated plasmons are collective excitations of correlated electrons that arise in strongly correlated electron systems arising from substantial long-range electronic correlations (non-local Coulomb interactions). Having confirmed the presence of correlated plasmons in the superconducting LSCO, it is necessary to assess if the long-range electronic correlations leading to the formation of correlated plasmons has a direct role in mediating superconductivity. It has previously been reported that long-range electronic correlations are not only present in the ground states, but also play a direct role in the collective excitation modes of correlated materials [42]. Previous work showed that a suitable magnitude of long-range electronic correlations strength give rise to a pair of plasmonic features with two-fold energy relation (at $U^*$ and $U^*/2$, where long-range electronic correlations have been accounted for in the $U^*$ component) in the LF spectra [40]. The feature at $U^*$ appears concomitantly with collective excitations between the Mott-correlated bands, while that at position $U^*/2$ can be attributed to particle-hole excitations between the Hubbard band and the quasiparticle peak at $E_F$ [39]. Based on a comprehensive analysis of long-range electronic correlations supported by our many-body numerical calculations and experimentally-derived LF spectra (Figures 2-3), feature $A^*$ (~1.84 eV) in the LF spectra of superconducting LSCO/LSAO can be identified to be equivalent to $U^*$, while feature $A$ (~0.92 eV), whose energy scale is half of the feature $A^*$ can be ascribed to $U^*/2$. Similarly, the two-fold energy relation of the peaks in the LF spectra of superconducting LSCO films on other substrates can also be attributed to the $U^*$ and $U^*/2$ band features unique to the correlated plasmons.

The plasmon peak profile of the respective superconducting LSCO films are summarized in Figure 4a. Features $A$ and $A^*$ always maintain a two-fold relationship that is consistent with $U^*$ and $U^*/2$. Conversely, LF spectra of the non-superconducting LSCO films do not display these features. This may point to the absence of long-range electronic correlations in these non-superconducting films. Our observations may reveal the possibility that the in-plane long-range electronic correlations may play an important role in mediating superconductivity. In addition, the relation of energy position of plasmon $A^*$ ($U^*$) and in-plane interfacial strain (Table. S6) is plotted in Figure. 4b. The energy position of plasmon $A^*$ increases with increasing in-plane interfacial tensile strains. However, as the in-plane interfacial tensile strains continues to increase over 0.34%, the energy position of plasmon $A^*$ decreases in reverse. The trend suggests that interfacial strain has influence on the effective long-range electronic correlations and there exists an optimal value for the modulation.

## 2.4 Theoretical calculation of extended Hubbard model

To provide further insight on the effect of the long-range electronic correlations, numerical simulations are conducted for the two-dimensional (2D) extended Hubbard model by employing dynamical cluster approximation (DCA) with a continuous-time auxiliary-field (CT-AUX) quantum Monte Carlo (QMC) technique. We note that, although the well-known charge-transfer insulating nature of undoped cuprates ideally calls for a more sophisticated theoretical treatment based on multi-orbital Hubbard models that explicitly include oxygen degrees of freedom—a task that remains computationally formidable—our single-orbital model nonetheless supports our conclusions regarding the critical role of long-range electronic correlations [43]. In particular, this many-body calculation examines the impact of the longer-range next-nearest-neighbor Coulomb repulsion $V'$ in addition to the nearest-neighbor repulsion $V$ by employing the extended Hubbard model (Eq. S13). To compare the calculation results with the experimentally-acquired LF data, we extract the local density of states (DOS) by employing the maximum entropy method (MEM) of analytical continuation applied to the self-energy in Matsubara frequency calculated from the DCA [44, 45].

Figure 4c compares the DOS of the simulation that consists of both $V$ and $V'$ (solid lines) and that with only $V$ (dashed lines). Generally, the presence of $V'$ regularizes the DOS of the curves that only consists of $V$. For instance, under the condition of $V/t=0.8$ (blue line), some insignificant features will vanish in the presence of $V'$. Besides, the DOS peaks become more prominent when $V'\neq 0$, such as in the case where $V/t=1.0$ (see $\omega_+$, $\omega_-$ peaks of red line). More importantly, by incorporating the $V'$ parameter into the calculation, the asymmetry (dashed spectra) around the point $\omega=0$ can be rectified as indicated by the spectra in solid lines. The electronic transitions between the three peaks (denoted as $\omega_-$, $\omega_0$, and $\omega_+$) of the DOS correspond to the $A$ and $A^*$ peaks of the experimental LF spectra (Figure 3), notably with the energy of feature $A^*$ located at the energy position that is twice of feature $A$. The symmetry over $\omega=0$ of the DOS may suggest that the deduced peaks could be consistent with features $A$ & $A^*$ of the LF spectra.

To better analyze the DOS under the condition where $V'\neq 0$, we define $\Delta A\equiv|\omega_++\omega_-|$, $\Delta B\equiv|\omega_+-\omega_-|$ to characterize the asymmetry of the DOS and peak distance corresponding to the energy position of feature $A^*$ in the experimental LF spectra respectively. As displayed in Figure 4d, the evolution of $\Delta A$ and $\Delta B$ versus $V/t$ both display a nontrivial dip of the peak structure. This result suggests that there exists an 'optimal' value for $V/t$ that governs the long-range electronic correlations. Specifically, the dip in $\Delta A$ implies that the optimal intermediate $V/t\sim 1.0$ results in the most symmetric DOS. Meanwhile, the peak feature of $\Delta B$ at $V/t\sim 1.0$ may be closely tied to the optimal scale of long-range electronic correlations under interfacial strain as discussed in Figure 4b. As a supplementary observation, Figure 4e could be seen as providing further support for the importance of long-range interactions. The sole presence of the local Hubbard $U$ does not lead to any non-trivial dependence of either the symmetry or the peak position difference in DOS on the single parameter $U$. This theoretical analysis suggests that long-range electronic correlations could be an intrinsically important factor in accounting for the physical properties of copper oxides, in alignment with our experimental observations. Despite these encouraging results, future work employing the multi-orbital model with explicit oxygen orbitals (with controlled approximations) is strongly desired to achieve a detailed and quantitative understanding of the charge excitations and collective modes associated with correlated plasmons. We also note that, rather than relying solely on the calculated DOS, plasmons are collective charge oscillations that are most rigorously identified via the dielectric function or charge susceptibility, which is, unfortunately, computationally much more demanding than the DOS and therefore falls outside the scope of the present work.

**2.5 The modulation of long-range electronic correlations on $T_c$**

Have confirmed the effect impact from long-range electronic correlations in our system through theoretical calculation and experimental evidence, we further notice that the $T_c$ varies with respect to the interfacial strain exacted on the LSCO films by different substrates and film thicknesses (Figure 4f). Interfacial strain analysis indicates that only the LSCO/LSAO films experience in-plane compressive strain, while LSCO films on the other substrates experience in-plane tensile strain (Table S6). Comparing the strain analysis with the corresponding LF spectra, it is inferred that long-range electronic correlations have an effect on the $T_c$ via the effects of strain in superconducting LSCO. Under the influence of in-plane compressive strain, LSCO/LSAO films register a higher $T_c$ compared to those on other substrates. This is due a reduction in in-plane atomic distance of the $CuO_2$ layer as the lattice is compressed, thereby enhancing the non-local Coulomb interactions (long-range electronic correlations) in the LSCO films. This in turn results in a spectral weight transfer of the plasmonic modes from feature $A^*$ ($U^*$) to $A$ ($U^*/2$). Such enhanced long-range electronic correlations directly influence the superconductive properties of the LSCO films and increase the $T_c$. On the contrary, with the increase in in-plane atomic distance of $CuO_2$ layer under in-plane tensile strain of LSCO/LAO, LSCO/STO, LSCO/LSAT samples, the non-local Coulomb interactions (long-range electronic correlations) weaken, thereby reducing their $T_c$.

However, the effects of a reduced long-range electronic correlations are negated in LSCO films under excessive in-plane tensile strain and this destroys the plasmon pair as seen in the absence of the plasmon pair in the ~9.0-nm LSCO film (Figures 3d-f). This coincidence of plasmon pair in the LF spectra and superconductivity in thicker LSCO films clearly indicates that long-range electronic correlations mediated by interfacial lattice strain indirectly mediate the onset of superconductivity and modulate the corresponding $T_c$ in high-temperature superconductors. Our systematic experimental verification provides a detailed understanding of the critical role of long-range electronic correlations in mediating superconductivity in high-temperature superconductors and offers new insights for the elucidating the high-temperature superconducting mechanism.

## 3. Conclusion

In conclusion, our present work has unveiled critical insights into high-energy excited long-range electronic correlations in high-temperature cuprate superconductors. Through a comprehensive series of experimental investigations combined with dynamical cluster approximation (DCA) quantum Monte Carlo calculation of the extended Hubbard model with long-range interactions, the observed plasmonic states and anomalous optical behavior are linked to strain- and substrate-dependent shifts in the electronic structure, mediated by high-energy excitations that modulate long-range electronic correlations. Notably, these trends indicate that interfacial lattice strain in LSCO films, induced by varying substrate materials, tunes long-range electronic correlations. Such correlations may indirectly contribute to the onset of superconductivity and affect the transition temperature in high-temperature superconductors. This work offers a more complete description of the high-energy optical response and clarifies intricate relationship between strain-excited long-range electronic correlations and the emergence of superconductivity. The identification of high-energy collective excitations and long-range electronic correlations opens new avenues for understanding the fundamental mechanisms of high-temperature superconductivity. Moreover, these findings are important stepping stones in the exploration of quasiparticle excitations and optical coupling in other strongly correlated electron systems, with potential implications for future superconducting materials and device applications.

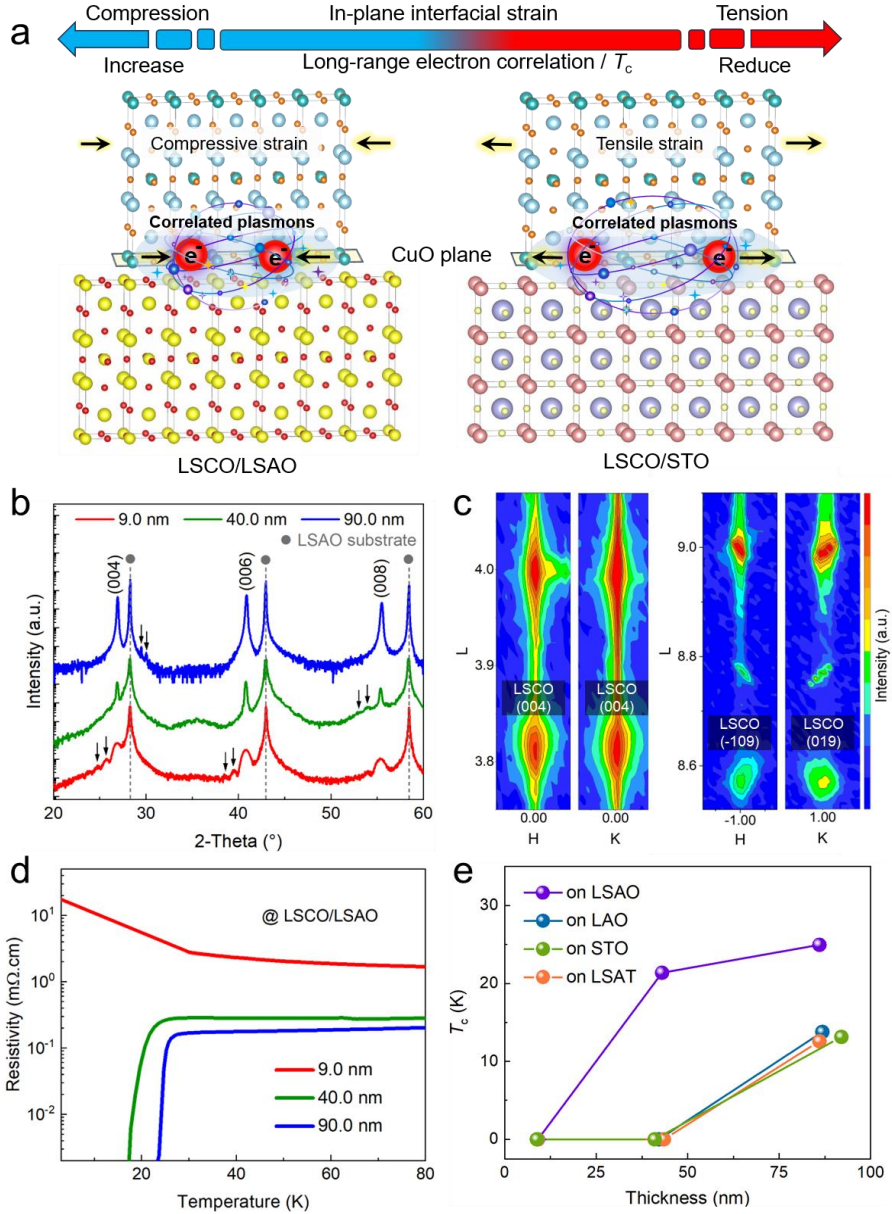

**Figure 1.** The long-range correlated plasmon dynamics in the LSCO layer which has been confirmed to be of good crystalline quality by XRD characterization and transport measurements. (a) Schematic depicting the LSCO/LSAO and LSCO/STO crystal structure as well as the underlying mechanism of correlated plasmons modulated by interfacial strains. In the LSCO/LSAO film, an in-plane compressive strain reduces the atomic distance, which enhances long-range electronic correlations, leading to an increase in $T_c$. In contrast, the LSCO/STO film experiences in-plane tensile strain, increasing the atomic distance, which diminishes long-range electronic correlations and results in a decrease in $T_c$. (b) XRD pattens about the LSCO/LSAO (00$l$) peaks with different thickness. (c) RSMs around the $(004)_{HL}$, $(004)_{KL}$, $(-109)_{HL}$ and $(019)_{KL}$ Bragg reflection for LSCO/LSAO films (d) Temperature-dependent resistivity of LSCO/LSAO films with different thicknesses. (e) A summary of $T_c$ for LSCO/LSAO, LSCO/LAO, LSCO/STO, LSAT films of different thicknesses.

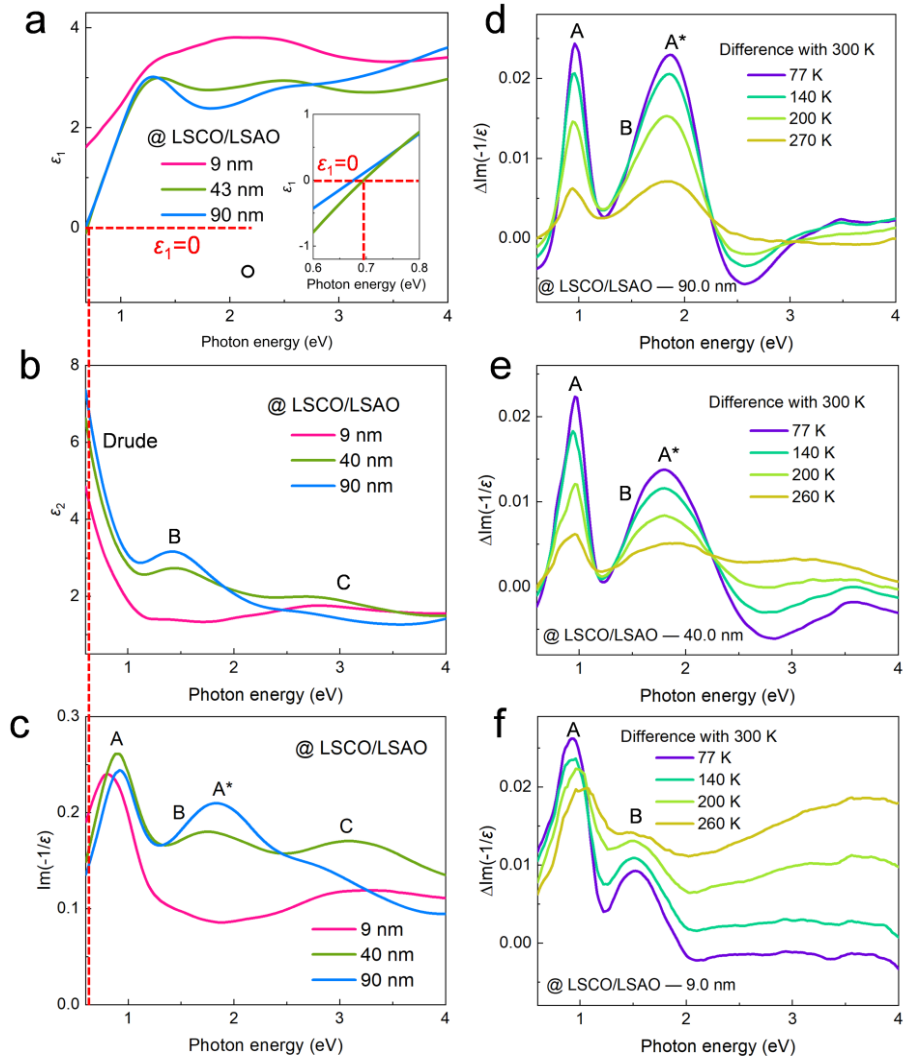

**Figure 2**. Plasmons detection in LSCO/LSAO system at 300 K. (a) The real ($\varepsilon_1$) and (b) imaginary ($\varepsilon_2$) of complex dielectric function ($\varepsilon(\omega)$), the inset of (a) is enlarged zero-crossing of $\varepsilon_1$. (c) The energy loss function (LF) spectral (Im(-1/$\varepsilon(\omega)$)) of LSCO/LSAO films with different thickness. The differential spectra of the LF idented as for LSCO/LAO film of (d) ~90.0 nm, (e) ~40.0 nm and (f) ~9.0 nm.

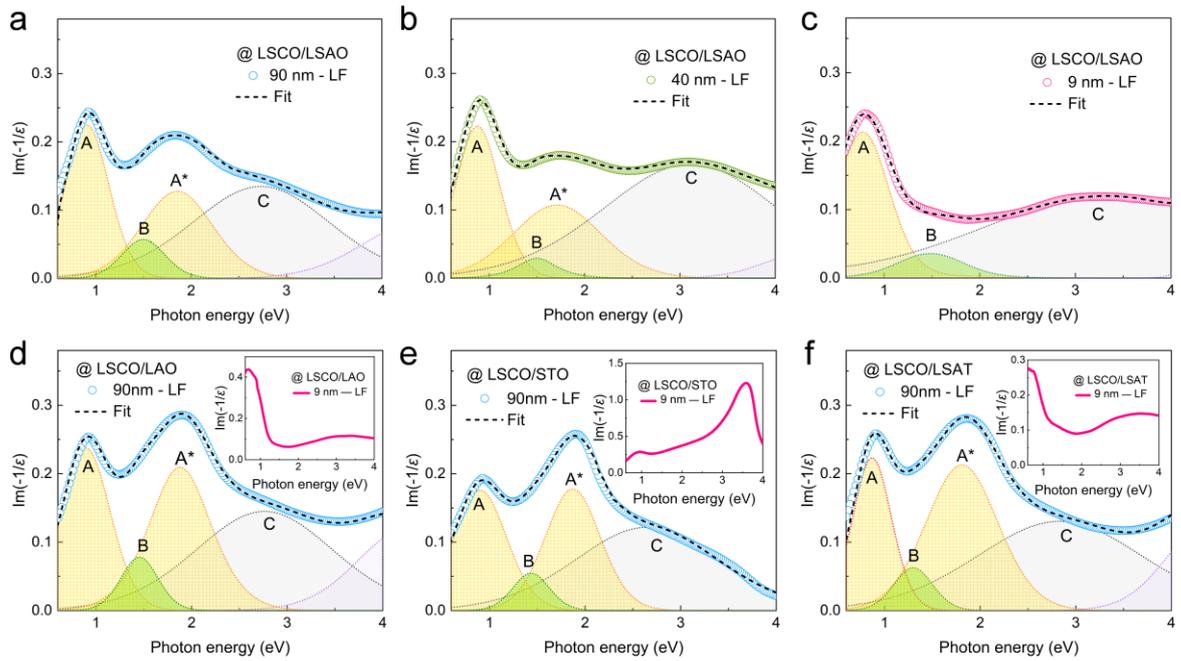

**Figure 3.** LF spectral and respective fitting features of LSCO films with different thickness on varying substrates. The open circle curve is measured experimental LF data, the black dashed line through the data represents the total fit. LF spectral and respective fitting curves of LSCO/LSAO films for (a) ~90.0-nm, (b) ~40.0-nm and (c) ~9.0-nm. LF spectral and respective fitting curves of (d) ~90.0 nm LSCO/LAO film, (e) ~90.0 nm LSCO/STO film, (f) ~90.0 nm LSCO/LSAT film. Inset is their respective LF spectral for ~9.0-nm non-superconducting compound.

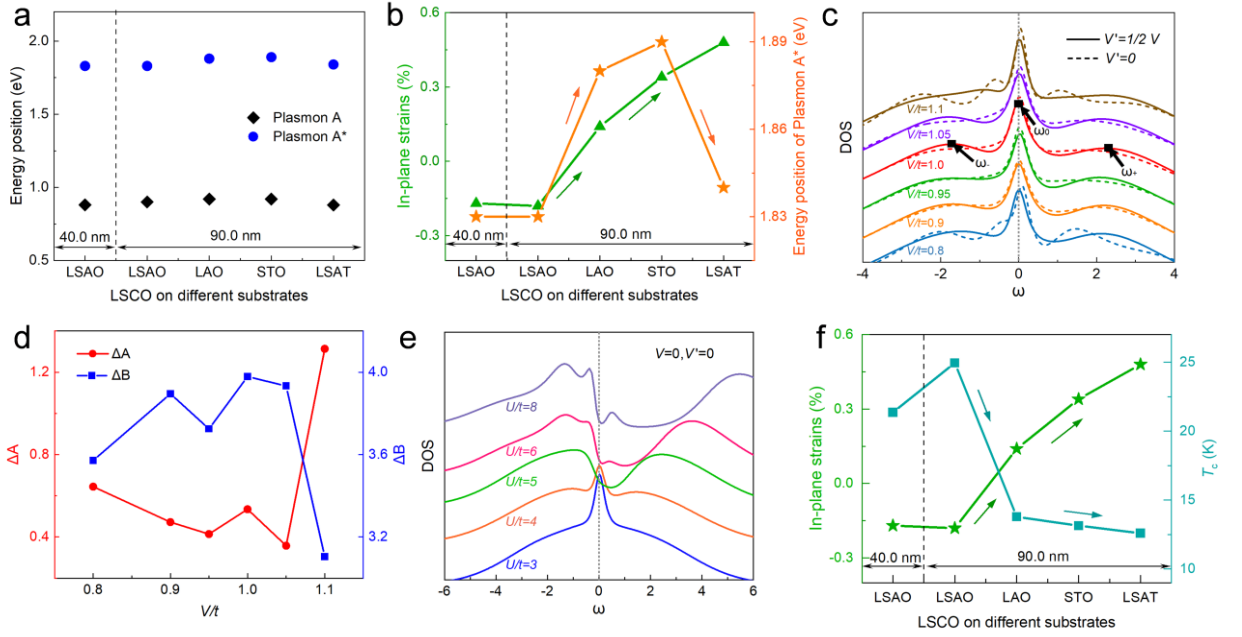

**Figure 4.** The trend in the long-range electronic correlations and DOS generated by the extended Hubbard model. (a) Energy position of plasmons *A* and *A*$^*$ for the superconducting LSCO films. (b) The energy position of plasmon *A*$^*$ and the in-plane interfacial strain as a function of film thickness caused by the respective substrates. (c) Variations in DOS structures under different conditions with the presence of *V* and *V'* (solid spectra) and that with only *V* (dashed lines); (d) The extracted asymmetry parameter Δ*A* and the peak distance Δ*B* corresponding to *A*$^*$ peak of experimental LF; (E) The Hubbard model results with only Hubbard *U* in the absence of *V* and *V'*. The parameters are $N_c$=4, *U*=4*t*, *T*=0.1*t* and *n*=0.98. (F) Critical temperatures and in-plane interfacial strain as a function of film thickness on various substrates.

**Supporting Information**
The information for discussion on sample preparation (PLD method), crystalline quality characterizations (XRD, RSM, R-T), spectroscopic ellipsometry measurements (principle, methods, model, and data analysis), analysis on interfacial strains, details of DCA calculations of the extended Hubbard model and supplementary figures and associated references are provided in the Supplementary Materials.

**Acknowledgements**
We thank G. A. Sawatzky (Stewart Blusson Quantum Matter Institute, University of British Columbia) for stimulating discussion on the interpretation and modeling of the experimental findings. This work was supported by National Natural Science Foundation of China (Grant Nos. 12374378, 52172271, 52307026, 52477022), the National Key R&D Program of China (Grant No. 2022YFE03150200), Shanghai Science and Technology Innovation Program (Grant No. 23511101600), the Ministry of Education, Singapore, under MOE-T2EP50124-0003. Xiongfang Liu acknowledges the support from the China Scholarship Council (CSC) (Grant No. 202306890069). Chi Sin Tang acknowledges the support from the NUS Emerging Scientist Fellowship. Xun Liu and Mi Jiang acknowledge the support from National Natural Science Foundation of China (Grant No. 12174278) and Priority Academic Program Development of Jiangsu Higher Education Institutions.

**Conflict of Interest**
The authors declare no conflict of interest.

**Data Availability Statement**
The data that support the findings of this study are available from the corresponding authors upon reasonable request.

# REFERENCES

1. B. C. D. Lee, Y. Shi, G.-Y. Kim, N. Campbell, F. Xue, K. Song, S.-Y. Choi, J. P. Podkaminer, T. H. Kim, P. J. Ryan, J.-W. Kim, T. R. Paudel, J.-H. Kang, J. W. Spinuzzi, D. A. Tenne, E. Y. Tsymbal, M. S. Rzchowski, L. Q. Chen, J. Lee, C. B. Eom Isostructural metal-insulator transition in VO2. *Science* **362,** 1037-1040 (2018).
2. J. Inoue&S. Maekawa Spiral State and Giant Magnetoresistance in Perovskite Mn Oxides. *Phys Rev Lett* **74,** 3407-3410 (1995).
3. A. Garg, M. Randeria&N. Trivedi Strong correlations make high-temperature superconductors robust against disorder. *Nature Physics* **4,** 762-765 (2008).
4. L. Zhang, et al. The Mechanism of Electrolyte Gating on High-T(c) Cuprates: The Role of Oxygen Migration and Electrostatics. *ACS Nano* **11,** 9950-9956 (2017).
5. L. F. Valentin Crépel New mechanism and exact theory of superconductivity from strong repulsive interaction. *Sci. Adv.* **7,** eabh2233 (2021).
6. T. E. Mason, G. Aeppli, S. M. Hayden, A. P. Ramirez&H. A. Mook Low energy excitations in superconducting La1.86Sr0.14CuO4. *Phys Rev Lett* **71,** 919-922 (1993).
7. H. A. M. S. M. Hayden, Pengcheng Dai, T. G. Perring, F. Dogˇan The structure of the high-energy spin excitations in a high-transitiontemperature superconductor. *NATURE* **429,** 531-534 (2004).
8. X. Yin, et al. Unraveling how electronic and spin structures control macroscopic properties of manganite ultra-thin films. *NPG Asia Materials* **7,** e196-e196 (2015).
9. I. F. Herbut Effective theory of high-temperature superconductors. *Phys Rev Lett* **94,** 237001 (2005).
10. X. Yin, et al. Coexistence of Midgap Antiferromagnetic and Mott States in Undoped, Hole- and Electron-Doped Ambipolar Cuprates. *Phys Rev Lett* **116,** 197002 (2016).
11. J. Z. Shusen Ye, Zhiheng Yao, Sixuan Chen, Zehao Dong, Xintong Li, Luchuan Shi, Qingqing Liu, Changqing Jin, Yayu Wang Visualizing the Zhang-Rice singlet, molecular orbitals and pair formation in cuprate. *Arvix* 2309.09260 (2023).
12. B. Vignolle, et al. Two energy scales in the spin excitations of the high-temperature superconductor La2−xSrxCuO4. *Nature Physics* **3,** 163-167 (2007).
13. H. Yan, et al. Ubiquitous coexisting electron-mode couplings in high-temperature cuprate superconductors. *PNAS* **120,** e2219491120 (2023).
14. P. W. Anderson The Resonating Valence Bond State in La2CuO4and Superconductivity. *Science* **235,** 1196-1198 (1987).
15. Z. Z. G. Baskaran, P.W. Anderson The resonating valence bond state and high-Tc superconductivity—A mean field theory. *Solid State Communications* **63,** 973-976 (1987).
16. V. J. Emery Theory of high-Tc superconductivity in oxides. *Phys Rev Lett* **58,** 2794-2797 (1987).
17. M. Qin, T. Schäfer, S. Andergassen, P. Corboz&E. Gull The Hubbard Model: A Computational Perspective. *Annual Review of Condensed Matter Physics* **13,** 275-302

(2022).
18. A. Reymbaut, et al. Antagonistic effects of nearest-neighbor repulsion on the superconducting pairing dynamics in the doped Mott insulator regime. *Physical Review B* **94,** 155146 (2016).
19. H. Terletska, T. Chen, J. Paki&E. Gull Charge ordering and nonlocal correlations in the doped extended Hubbard model. *Physical Review B* **97,** 115117 (2018).
20. M. Jiang, U. R. Hähner, T. C. Schulthess&T. A. Maier d-wave superconductivity in the presence of nearest-neighbor Coulomb repulsion. *Physical Review B* **97,** 184507 (2018).
21. J. Paki, H. Terletska, S. Iskakov&E. Gull Charge order and antiferromagnetism in the extended Hubbard model. *Physical Review B* **99,** 245146 (2019).
22. N. Plonka, C. J. Jia, Y. Wang, B. Moritz&T. P. Devereaux Fidelity study of superconductivity in extended Hubbard models. *Physical Review B* **92,** 024503 (2015).
23. M. Jiang Enhancing d-wave superconductivity with nearest-neighbor attraction in the extended Hubbard model. *Physical Review B* **105,** 024510 (2022).
24. U. G. Hau H. Wang, R. J. Thorn, K. Douglas Carlson, Mark A. Beno, Marilyn R. Monaghan, Thomas J. Allen, Roger B. Proksch, Dan L. Stupka Synthesis, structure, and superconductivity of single crystals of high-Tc La1.85Sr0.15CuO4, a lanthanum strontium copper oxide. *Inorganic Chemistry* **26,** 1190-1192 (1987).
25. L. W. QIN Yue-ling, DONG Xiao-li, ZHAO Bai-ru Growth Mode of Superconducting La2xSrxCu1+yO4 Thin Films on LaAlO3 Substrate. *Chinese Physics Letters* **5,** 530-532 (1998).
26. S. W. Zeng, et al. Observation of perfect diamagnetism and interfacial effect on the electronic structures in infinite layer Nd(0.8)Sr(0.2)NiO(2) superconductors. *Nat Commun* **13,** 743 (2022).
27. M. Kawai, F. Nabeshima&A. Maeda1 Transport properties of FeSe epitaxial thin films under in-plane strain. *Journal of Physics: Conf. Series* **1054,** 012023 (2018).
28. Y. Dong, et al. Interfacial Octahedral Manipulation Imparts Hysteresis‐Free Metal to Insulator Transition in Ultrathin Nickelate Heterostructure. *Advanced Materials Interfaces* **6,** 1900644 (2019).
29. M. N. H. Sato Increase in the superconducting transition temperature by anisotropic strain effect in (001) La1.85Sr0.15CuO4 thin films on LaSrA10 4 substrates. *Physica C* **274,** 221-226 (1997).
30. T. L. Meyer, L. Jiang, S. Park, T. Egami&H. N. Lee Strain-relaxation and critical thickness of epitaxial La1.85Sr0.15CuO4 films. *APL Materials* **3,** 126102 (2015).
31. C. C. Chen, et al. Doping evolution of the oxygenK-edge x-ray absorption spectra of cuprate superconductors using a three-orbital Hubbard model. *Physical Review B* **87,** 165144 (2013).
32. J. Zaanen, G. A. Sawatzky&J. W. Allen Band gaps and electronic structure of transition-metal compounds. *Phys Rev Lett* **55,** 418-421 (1985).
33. M. A. Naradipa, P. E. Trevisanutto, T. C. Asmara, M. A. Majidi&A. Rusydi Role of hybridization and on-site correlations in generating plasmons in strongly correlated

La2CuO4. *Physical Review B* **101,** 201102(R) (2020).
34. S. Uchida, et al. Optical spectra of La2-xSrxCuO4: Effect of carrier doping on the electronic structure of the CuO2 plane. *Phys Rev B Condens Matter* **43,** 7942-7954 (1991).
35. N. Nucker, et al. Plasmons and interband transitions in Bi2Sr2CaCu2O8. *Phys Rev B Condens Matter* **39,** 12379-12382 (1989).
36. X. Yin, et al. Quantum Correlated Plasmons and Their Tunability in Undoped and Doped Mott-Insulator Cuprates. *ACS Photonics* **6,** 3281-3289 (2019).
37. S. A. Maier *Plasmonics: Fundamentals and Applications* Springer Science & Business Media (2007).
38. J. B. Kana Kana, G. Vignaud, A. Gibaud&M. Maaza Thermally driven sign switch of static dielectric constant of VO 2 thin film. *Optical Materials* **54,** 165-169 (2016).
39. E. G. van Loon, H. Hafermann, A. I. Lichtenstein, A. N. Rubtsov&M. I. Katsnelson Plasmons in strongly correlated systems: spectral weight transfer and renormalized dispersion. *Phys Rev Lett* **113,** 246407 (2014).
40. T. Ayral, P. Werner&S. Biermann Spectral properties of correlated materials: local vertex and nonlocal two-particle correlations from combined GW and dynamical mean field theory. *Phys Rev Lett* **109,** 226401 (2012).
41. M. Sun, et al. Tunable Collective Excitations in Epitaxial Perovskite Nickelates. *ACS Photonics* **11,** 2324-2334 (2024).
42. P. Hansmann, T. Ayral, L. Vaugier, P. Werner&S. Biermann Long-range Coulomb interactions in surface systems: a first-principles description within self-consistently combined GW and dynamical mean-field theory. *Phys Rev Lett* **110,** 166401 (2013).
43. E. Gull, P. Werner, O. Parcollet&M. Troyer Continuous-time auxiliary-field Monte Carlo for quantum impurity models. *EPL (Europhysics Letters)* **82,** 57003 (2008).
44. J. E. G. Mark Jarrell Bayesian inference and the analytic continuation of imaginary-time quantum Monte Carlo data. *Physics Reports* **269,** 133-195 (1996).
45. M. J. Thomas Maier, Thomas Pruschke, Matthias H. Hettler Quantum cluster theories. *REVIEWS OF MODERN PHYSICS* **77,** 1027-1080 (2005).